\documentclass[lettersize,journal]{IEEEtran}
\usepackage{amsmath,amsfonts}
\usepackage{algorithmic}
\usepackage{algorithm}
\usepackage{array}
\usepackage[caption=false,font=normalsize,labelfont=sf,textfont=sf]{subfig}
\usepackage{textcomp}
\usepackage{stfloats}
\usepackage{url}
\usepackage{verbatim}
\usepackage{graphicx}
\usepackage{cite}
\usepackage{xcolor}
\usepackage[colorlinks,linkcolor=black,anchorcolor=black,citecolor=black]{hyperref}
\graphicspath{{fig/}}
\begin{document}

\title{Wireless Imaging for Low-Altitude Surveillance: \\A New Paradigm for ISAC Networks}

\author{Yixuan Huang, Jie Yang, Chao-Kai Wen, \IEEEmembership{Fellow, IEEE}, Shi Jin, \IEEEmembership{Fellow, IEEE}

\thanks{Yixuan Huang, Jie Yang, and Shi Jin (corresponding author) are with Southeast University, China;
Chao-Kai Wen is with the Institute of Communications Engineering, National Sun Yat-sen University, Taiwan.}}

\maketitle

\begin{abstract}
The rapid growth of the low-altitude economy calls for integrated sensing and communication (ISAC) networks capable of robust flight monitoring.
This article advocates wireless imaging as a unifying sensing paradigm that enables ISAC networks to function as comprehensive low-altitude guardians.
We present a hierarchical imaging framework that enhances sensing capability from wide-area snapshot imaging to dynamic trajectory-aware imaging and target-centric fine-grained characterization.
At the wide-area level, low-altitude sensing is reformulated as a spatial imaging problem, where distributed base stations and communication users collaboratively construct a holistic view of the aerial space.
Building on this foundation, multi-frame dynamic imaging exploits temporal correlations to support robust trajectory tracking and prediction under mobility-induced challenges such as occlusions.
For security-critical scenarios, the framework further enables fine-grained imaging of flying and hovering uncrewed aerial vehicles, providing detailed characterization beyond conventional point-target abstractions.
Additionally, we propose a novel evaluation metric named imaging coverage to examine the sensing fidelity of the proposed framework.
Illustrative case studies demonstrate the potential of imaging in ISAC networks to support wide-area monitoring, motion-aware tracking, and fine-grained target analysis.
\end{abstract}

\section{Introduction}
\label{sec-intro}

The rapid expansion of the low-altitude economy is reshaping how airspace is utilized, enabling emerging applications such as aerial logistics and intelligent traffic monitoring \cite{wu2025toward}.
As uncrewed aerial vehicles (UAVs) become increasingly pervasive and autonomous, low-altitude airspace is evolving into a densely occupied and highly dynamic environment.
This evolution calls for continuous and reliable airspace awareness to ensure safe and efficient UAV operations \cite{wang2025toward}.

Conventional low-altitude surveillance primarily relies on dedicated sensing infrastructures, including satellite systems, optical cameras, and radars.
Although effective, these solutions introduce additional hardware cost, deployment complexity, and operational constraints, which may limit the scalability and economic viability of large-scale low-altitude services \cite{jin2025co}.
Integrated sensing and communication (ISAC) has therefore emerged as a promising alternative, enabling low-altitude sensing by reusing existing wide-area communication networks while supporting all-day and all-weather operations \cite{jiang2025integrated}.

Despite its potential, realizing reliable low-altitude surveillance faces multifaceted demands.
First, beyond isolated position estimates, low-altitude sensing must capture multi-target spatial relationships and environmental interactions to support cooperative monitoring and intrusion detection \cite{ma2024networked}.
Second, UAVs exhibit highly dynamic motion patterns, which require continuous sensing and timely trajectory awareness rather than snapshot-based detection \cite{yang2025cooperative}.
Third, in security-critical scenarios, fine-grained characteristics such as shape and size become indispensable for reliable target recognition (e.g., distinguishing UAVs from birds), where traditional point-target abstraction is physically invalid \cite{li2025uav}.
Confronted with these sensing requirements, conventional localization-centric approaches prove fundamentally insufficient, highlighting the need for a new sensing paradigm.

Recently, wireless imaging has attracted growing interest in ISAC systems as a new sensing paradigm.
In this context, wireless imaging is defined as the process of reconstructing the spatial scattering information and 3D geometry of a region of interest (ROI) directly from channel state information (CSI) measurements \cite{luo2024integrated}.
Unlike visible-light imaging, wireless imaging is inherently robust to lighting and weather conditions.
More importantly, wireless imaging directly yields an explicit and holistic representation of the sensed environment, avoiding the cascaded processing and ambiguity accumulation commonly encountered in localization-based approaches \cite{huang2025learned}.

Motivated by these advantages, this article advocates wireless imaging as a unifying low-altitude surveillance framework for ISAC networks.
Specifically, CSI measurements collected by distributed network nodes can be fused at the data level, enabling both wide-area airspace monitoring and target-centric fine-grained sensing that goes beyond point-target abstractions \cite{wang2025dreamer}.
While low-altitude imaging is inherently challenging due to sparse measurements and the ill-posed nature, emerging techniques like artificial intelligence (AI) offer powerful tools to integrate data-driven learning with physical constraints, rendering high-fidelity reconstruction feasible \cite{luo2025airguard}.

\begin{figure*}[t]
\centering
\includegraphics[width=0.9\linewidth]{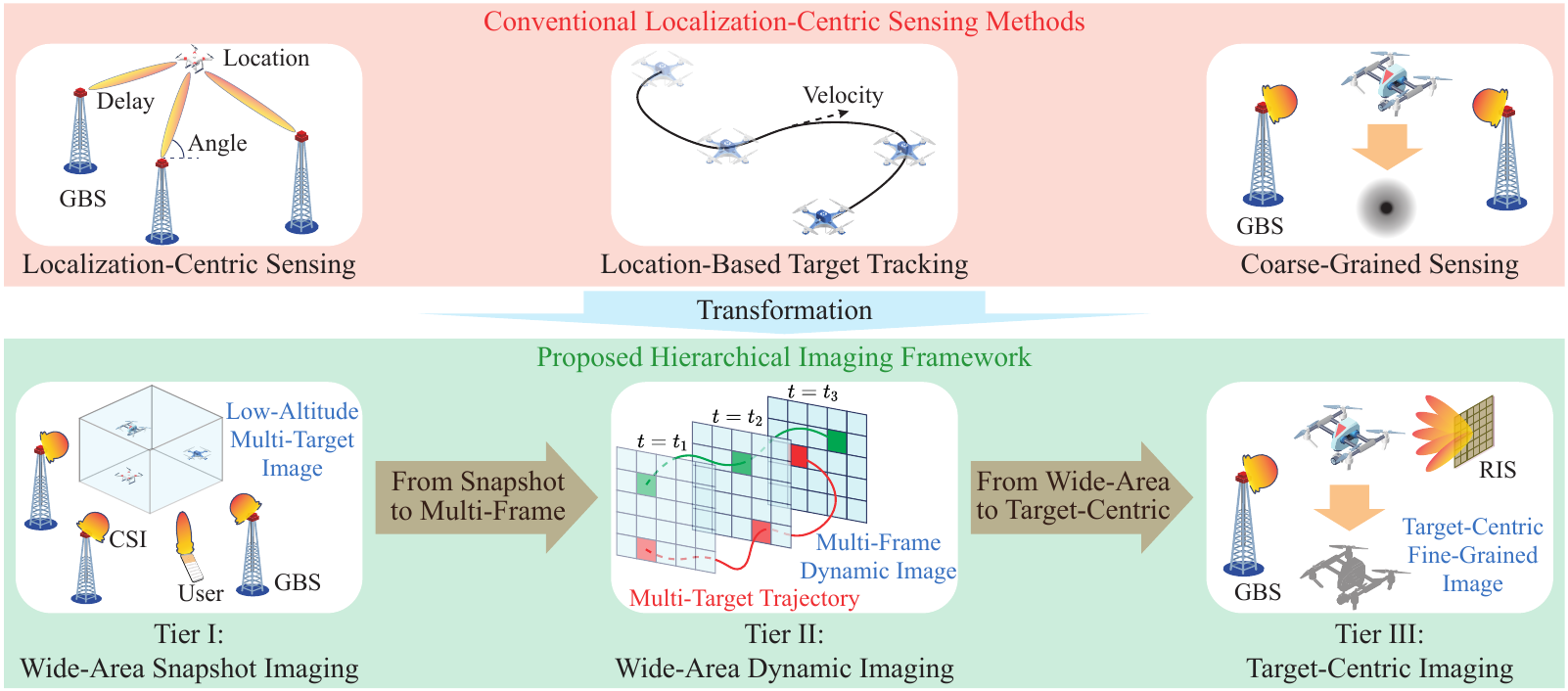}
\vspace{-0.3cm}
\caption{Paradigm shift from traditional schemes to the proposed wireless imaging-based framework.}
\vspace{-0.3cm}
\label{fig1}
\end{figure*}

Building on this vision, we present an imaging-based hierarchical low-altitude sensing framework that explicitly transforms conventional sensing paradigms through a three-tiered progression, as illustrated in Fig.~\ref{fig1}.
First, conventional localization-centric sensing is elevated to wide-area imaging, enabling holistic visualization of low-altitude airspace with multiple targets.
Second, snapshot-based sensing is extended to multi-frame dynamic imaging, allowing continuous capture of UAV motion patterns and trajectory evolution.
Third, the framework supports a transition from wide-area monitoring to target-centric fine-grained imaging, revealing detailed shape and size characteristics for security-critical analysis.
Through these transformations, detection, localization, tracking, and imaging are naturally unified within a common framework.

The contributions of this work are summarized as follows:
\begin{itemize}
\item We present an imaging-based wide-area low-altitude sensing framework in ISAC networks, whose sensing coverage is enhanced leveraging existing infrastructures.
\item We extend the paradigm from snapshot sensing to multi-frame dynamic imaging, enabling continuous tracking and prediction of UAV kinematics.
\item We investigate target-centric imaging of intrusive UAVs, which formulates fine-grained images with shape and size information to facilitate security risk assessment.
\end{itemize}

The rest of this paper is structured as follows:
Secs. \ref{sec-large-scale-imaging}, \ref{sec-dynamic-imaging}, and \ref{sec-fine-grained-imaging} detail the framework's three tiers: wide-area snapshot imaging, wide-area dynamic imaging, and target-centric imaging, respectively.
Sec. \ref{sec-results} presents illustrative simulation results, and Sec. \ref{sec-conclusion} concludes the paper and lists future directions.
Next, we begin with wide-area snapshot imaging in Tier I, as it lays the foundation for the subsequent dynamic and target-centric imaging.

\section{Snapshot Imaging for Wide-Area Low-Altitude Awareness}
\label{sec-large-scale-imaging}

This section details the first tier of the proposed hierarchical sensing framework, where conventional coordinate-based localization is shifted to comprehensive low-altitude imaging, as depicted in Fig. \ref{fig1}.
To achieve ubiquitous awareness, we propose extending imaging coverage through collaborative ISAC networks, synergizing ground base stations (GBSs), cooperative users, and new network elements (NNEs).

\subsection{Snapshot Sensing: From Localization to Imaging}
\label{sec-large-scale-imaging-1}

\begin{figure}[t]
\centering
\includegraphics[width=0.9\linewidth]{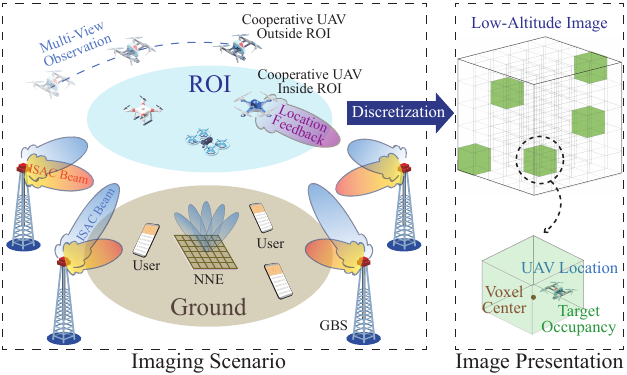}
\vspace{-0.3cm}
\caption{Wide-area snapshot imaging within ISAC networks.}
\vspace{-0.3cm}
\label{fig2}
\end{figure}

While target localization has long served as the cornerstone of low-altitude surveillance, it is increasingly becoming a bottleneck due to its inherent precision limitations and misalignment with emerging sensing demands, which include:
\begin{itemize}
\item \textbf{Scalability Limitations of Localization-Centric Sensing:} Traditional localization relies on step-wise parameter extraction, which lacks holistic environmental awareness and is prone to cascading errors. Moreover, its computational complexity surges with target density, making it ill-suited for future high-density low-altitude traffic.
\item \textbf{Mismatch Between Localization Outputs and Control Needs:} For safety-critical flight control, pinpointing an exact coordinate is often less valuable than defining spatial occupancy. The priority shifts from ``where is the point'' to ``which region is occupied'' to ensure safe separation and collision avoidance.
\item \textbf{Lack of Scattering-Aware Representation:} Beyond mere positions, scattering signatures are pivotal for target characterization, which serves as a capability essential for the tracking and security scenarios discussed later. Localization fails to capture this rich feature set.
\end{itemize}
Collectively, these deficiencies render traditional methods inadequate, necessitating novel low-altitude sensing paradigms.

Drawing inspiration from wireless imaging, we propose a novel low-altitude sensing framework.
As shown in Fig. \ref{fig2}, the airspace is conceptualized as a discretized volumetric image, where each voxel encodes the target's status via its scattering coefficient.
Unlike conventional methods, this image offers a unified representation, which simultaneously captures individual attributes (position, existence, and scattering coefficient) and the global topology, thereby providing comprehensive data for flight control and characterization.
Consequently, the sensing objective of the ISAC network undergoes a fundamental shift: moving from the explicit coordinate estimation of point targets to voxel-based imaging, which prioritizes the detection of spatial occupancy.
Crucially, this represents a direct mapping from CSI to the global scattering image, with its computational complexity decoupled from the number of targets.
Therefore, the imaging-based approach fundamentally resolves the bottlenecks inherent in traditional localization.

However, the transition to this imaging paradigm necessitates a comprehensive overhaul of underlying principles.
\textit{In terms of theoretical models}, the imaging-based framework demands high-fidelity channel modeling. Specifically, we must move beyond simplistic far-field assumptions to adopt near-field channel models, which are essential for preventing rank deficiency and ensuring information recoverability in low-altitude contexts.
\textit{In terms of performance limits}, a new analytical framework for imaging quality must be established to quantify the fundamental imaging capabilities of ISAC networks and rigorously assess the holistic reconstruction fidelity. This requires bridging the gap between classical imaging analysis, which is characterized by the point spread function and diffraction resolution limit, with traditional localization-based metrics.
\textit{In terms of algorithm design}, the core challenge shifts from low-dimensional parameter estimation to high-dimensional, ill-posed signal recovery. Addressing this requires adapting classical radar imaging principles while integrating AI-driven techniques to handle the complexity of ISAC networks and the sparse nature of low-altitude images.

\subsection{Enhancing Imaging Coverage via Network Cooperation}
\label{sec-large-scale-imaging-2}

Within the proposed imaging-based framework, a large imaging aperture should be synthesized to cover the low-altitude space.
However, given the vastness of the airspace, relying solely on individual nodes is insufficient, and network cooperation is essential to expand the sensing horizon.
This subsection presents the scalability of the proposed method by synergizing GBSs, users, and NNEs to achieve ubiquitous imaging coverage, as depicted in Fig. \ref{fig2}.

\subsubsection{GBS-Assisted Cooperation}

GBSs serve as the cornerstone of the collaborative ISAC network.
However, transforming them into effective low-altitude imagers entails distinct challenges.
The first challenge lies in dual-functional beamforming, which requires illuminating the entire low-altitude volume while maintaining seamless terrestrial communication. This is particularly challenging given the inherent downtilt configuration of existing commercial GBSs designed for ground users.
The second challenge is the rigorous management of interference, which arises from self-interference within a single node's transceiver arrays and mutual interference caused by simultaneous transmissions across multiple GBSs.
Mitigating these interferences requires hardware-software synergy, rational coordination of time-frequency resources, and a holistic redesign of the ISAC frame structure.
Ultimately, a fundamental topological limitation remains: due to the large inter-station distances, the imaging aperture synthesized by GBSs is inherently expansive but sparse, necessitating the support of cooperative users and NNEs.

\subsubsection{User-Assisted Cooperation}

Cooperative aerial and terrestrial users represent widely distributed, yet underutilized assets for low-altitude imaging.
First, cooperative UAVs within the ROI can provide coarse position estimates for themselves and nearby targets by establishing communication links with GBSs. These estimates can form preliminary aerial image support, which assists in solving the inverse imaging problem and mitigates its ill-posedness.
Second, UAVs operating outside the ROI serve as mobile scanning nodes. While fulfilling primary duties (e.g., traffic monitoring), they fly around the ROI to provide multi-view sensing perspectives.
Third, the massive population of ground users can be crowdsourced to synthesize a high-density sampling aperture, effectively remedying the spatial sparsity of the traditional GBS network.
However, orchestrating this massive collaboration introduces significant hurdles, particularly regarding real-time CSI feedback overhead, user location errors, and network-wide clock synchronization.
Thus, dedicated user selection strategy should be designed, and the ISAC frame structure should be optimized to reuse existing CSI feedback and synchronization processes in standard communications.

\subsubsection{NNE-Assisted Cooperation}

Emerging NNEs, such as reconfigurable intelligent surfaces (RISs) and network-controlled repeaters (NCRs), serve as pivotal components of the collaborative ISAC network \cite{an2025emerging}.
Specifically, RISs can reshape the electromagnetic environment with high energy efficiency, while NCRs amplify and forward signals under precise beamforming control.
Ground-deployed NNEs can intercept GBS signals originally directed at terrestrial users and reflect (or amplify) them toward the aerial space, extending the sensing coverage without requiring drastic modifications to GBS downtilt configurations.
However, unlocking this potential presents a complex multi-objective optimization problem, which requires a delicate trade-off between communication quality and sensing precision, involving the joint design of RIS phase shifts and NCR beamforming weights.
Moreover, practical considerations regarding deployment costs, power supply strategies, and long-term maintenance must be addressed to ensure the viability of these heterogeneous nodes.

\textbf{Evaluation Metric:}
Finally, we propose a systematic performance metric termed ``imaging coverage'' to quantify the ISAC network's surveillance capability.
Analogous to the concept of communication coverage \cite{gan2024coverage}, we define imaging coverage as the maximum volumetric region where the reconstruction accuracy exceeds a requisite fidelity threshold for a given resolution.
The interdependence is explicit: for a fixed imaging resolution, i.e., a constant voxel size determined by application needs, expanding the surveillance volume linearly increases the image's dimension.
This exacerbates the ill-posedness of the inverse problem, eventually degrading the imaging accuracy, which is typically evaluated by normalized mean square error (NMSE).
Therefore, imaging coverage represents the volumetric boundary where the reconstruction accuracy decays to the minimum acceptable threshold.
Serving as a comprehensive indicator, this metric encapsulates the impact of node density and topology, providing a feedback mechanism to guide network deployment.

\section{Dynamic Imaging for Continuous Tracking and Prediction}
\label{sec-dynamic-imaging}

\begin{figure}[t]
\centering
\includegraphics[width=0.9\linewidth]{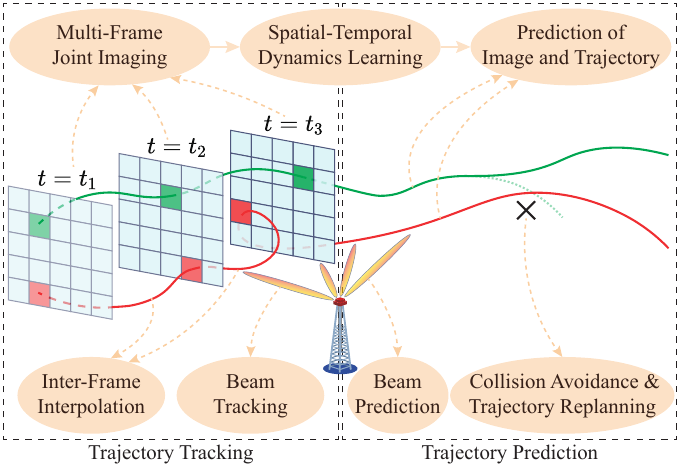}
\vspace{-0.3cm}
\caption{Dynamic imaging for trajectory tracking and prediction.}
\vspace{-0.3cm}
\label{fig4}
\end{figure}

Building upon the spatial imaging formulation established in the previous section, this section extends the hierarchical framework from static snapshots in Tier I to multi-frame dynamic imaging in Tier II.
Unlike traditional tracking schemes that are limited to single-target dynamics, our imaging-based approach inherently supports simultaneous multi-target tracking.
By capturing the dynamic evolution of the low-altitude topology over time, this method enables a holistic understanding of the airspace, thereby supporting robust, continuous trajectory tracking and prediction, as illustrated in Fig. \ref{fig4}.

\subsection{Why Temporal Imaging Matters}
\label{sec-dynamic-imaging-1}

Mobility is one of the defining characteristics of UAVs.
Consequently, the snapshot imaging paradigm discussed in the previous section is inherently insufficient in high-mobility scenarios.
To address this limitation, a transition from the purely spatial domain to a joint spatial-temporal sensing paradigm becomes inevitable.
Incorporating the temporal dimension unlocks two critical capabilities:

\subsubsection{Enhanced Surveillance via Temporal Correlation}

Driven by the continuous nature of UAV flight, low-altitude images exhibit strong inter-frame correlations. These correlations embed rich information regarding the UAV's kinematic features, which snapshot methods fail to capture, since they treat each instant in isolation.
In contrast, dynamic imaging unlocks this potential by integrating the temporal dimension.
By aggregating temporal redundancies and fusing them with velocity information, the framework generates time-evolving volumetric imagery. This allows for the robust reconstruction of UAV trajectories, effectively suppressing transient noise that plagues single-frame detection.
Moreover, this method enables super-resolution capabilities, where the learned spatio-temporal patterns can be leveraged to interpolate missing data in intermediate frames. Consequently, even under sparse sampling conditions, the framework ensures high-fidelity, continuous characterization of the target's presence and motion.

\subsubsection{Extension from Passive Tracking to Proactive Prediction}

Low-altitude networks demand the convergence of sensing, communication, computing, and control.
However, the rapid movement of UAVs leads to high sensitivity of real-time flight control due to communication and computation latencies.
To address this, dynamic imaging transcends the passive reconstruction of historical trajectories; it aims to extract UAV kinematic features to predict future states, thereby reducing latencies.
These two tasks are intrinsically coupled: while tracking involves solving the inverse imaging problem based on observed CSI, prediction focuses on learning the underlying causality and dynamics to generate future images.
Accurate trajectory prediction is pivotal, enabling precise beamforming and collision avoidance through timely trajectory replanning.
Furthermore, predicted trajectories can be fused with incoming CSI measurements, establishing a robust closed-loop prediction-update cycle, where the predicted results facilitate reliable state updates and mitigate severe estimation errors caused by intermittent or poor-quality measurements.

\subsection{Key Challenges and Enabling Techniques}
\label{sec-dynamic-imaging-2}

While multi-frame joint imaging promises significant performance gains, the inherent high mobility of UAVs poses unique challenges.

\subsubsection{Motion-Induced Defocusing in Multi-Frame Imaging}

The primary challenge in multi-frame integration is that UAV mobility inevitably introduces defocusing artifacts and ghosting across consecutive frames.
However, mathematically modeling these correlations is non-trivial, especially given the inherent sparsity of low-altitude images.
To overcome this, the ISAC system faces two critical demands. First, it must jointly process long-sequence CSI data to lock onto the spatial variations of non-zero voxels over time. Second, it must capture non-linear UAV motion patterns within a continuous time domain.
These demands position AI-driven sequence modeling as the core enabler.
The self-attention mechanism can guide the neural network (NN) to selectively focus on target-relevant features. Complementarily, neural ordinary differential equations can be employed to approximate state derivatives, facilitating robust kinematic feature learning.
Based on these AI-learned correlations, the system can realize effective motion compensation and UAV tracking, thereby resolving targets that were blurred in snapshot imaging.

\subsubsection{Occlusion-Aware Dynamic Imaging}

Target tracking in low-altitude environments is frequently disrupted by complex obstacles (e.g., buildings), creating scenarios of partial occlusion or discontinuous observation.
Traditional algorithms relying on frame-by-frame association often misinterpret temporary blockages as target ``deaths,'' creating fragmented trajectories.
To resolve this, the imaging framework requires a cognitive shift from reactive tracking to proactive continuity maintenance.
First, the system should distinguish temporary disappearance from target departure, requiring it to predict motion trends within the blind spots.
Second, upon the target's return, the system demands a global re-identification capability to match the new detection with the historical entity.
To address these issues, communication links can be exploited to mine kinematic motion features and assist in reconstructing trajectories within blind spots.
Moreover, space-time memory mechanisms can be employed to archive historical signatures (location, velocity, scattering) to create a ``memory bank,'' assisting in recognizing re-appearing targets.

\subsubsection{Latency-Fidelity Trade-off in Dynamic Imaging}

The dynamic imaging scheme faces a critical conflict: the trade-off between sensing fidelity and update latency.
While aggregating more historical frames suppresses noise and enhances resolution, it increases computational load and output delay, creating a bottleneck for resource-constrained edge devices.
Addressing this conflict requires a two-pronged strategy focused on adaptivity and efficiency.
On one hand, the system requires a dynamic mechanism to intelligently adjust the number of joint frames based on the target's current speed and CSI quality.
On the other hand, the underlying imaging algorithms must possess low computational complexity to ensure real-time responsiveness on the edge.
Emerging techniques provide the tools to achieve this equilibrium.
An adaptive time-window mechanism, potentially optimized via reinforcement learning, allows the system to autonomously balance accuracy and speed.
Moreover, the Mamba architecture \cite{zhang2025mamba} offers an alternative to realize linear-complexity fusion of historical data, paving the way for real-time dynamic perception.

\textbf{Refining the Metric for Dynamic Imaging:}
Finally, the ``imaging coverage'' metric established in snapshot imaging should be extended to the dynamic domain, evolving from defining a static detection zone to characterizing a continuous trackable volume.
This extended metric quantifies the temporal integration gain brought by multi-frame joint imaging, where coherent CSI data accumulation effectively lowers detection thresholds.
Consequently, this metric demonstrates how dynamic imaging expands coverage into previously invisible low-SNR regions, guiding network deployment to maximize not just the spatial area, but the continuity of trajectory tracking across the entire airspace.

\begin{figure*}[t]
\centering
\includegraphics[width=0.95\linewidth]{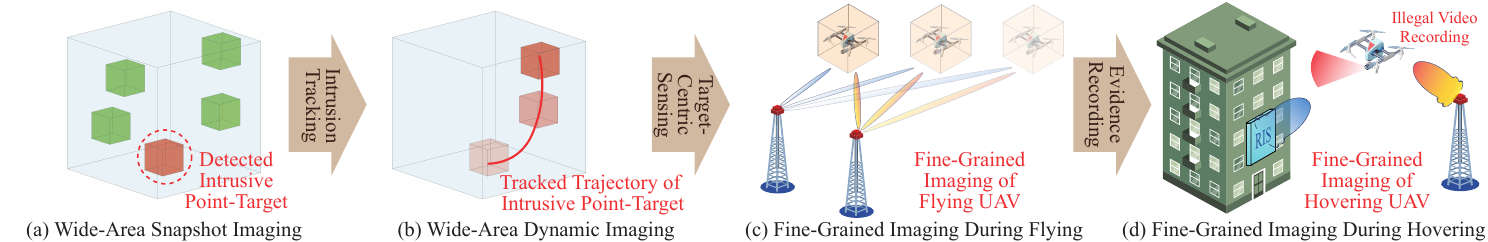}
\vspace{-0.3cm}
\caption{Fine-grained imaging and workflow of the proposed framework.}
\vspace{-0.3cm}
\label{fig5}
\end{figure*}

\section{Target-Centric Imaging for Security-Critical Scenarios}
\label{sec-fine-grained-imaging}

While wide-area imaging treats aerial targets as points in Tiers I and II, critical security scenarios necessitate a shift toward fine-grained target sensing in Tier III, as shown in Fig. \ref{fig5}.
Beyond mere localization, the system must characterize morphological attributes (e.g., shape and size) for risk assessment or capture high-fidelity evidence of illicit activities, such as unauthorized video recording.
This section investigates target-centric imaging tailored for these precision tasks.

\subsection{Motion-Aware Fine-Grained Imaging of Flying UAVs}
\label{sec-fine-grained-imaging-1}

By leveraging beam focusing, the ISAC network can constrict its ROI to the target's immediate vicinity, transitioning from wide-area detection to fine-grained morphological imaging.
Unlike the point-source modeling in Sec. \ref{sec-dynamic-imaging}, this approach aims to resolve the structural details of in-flight UAVs.
While inverse synthetic aperture radar (ISAR) provides the theoretical foundation for this task, its effective implementation is hindered by the complex flight dynamics of UAVs.
Specifically, UAVs exhibit three distinct motion modes, each posing unique challenges that necessitate tailored imaging strategies.

\subsubsection{Translational Motion Compensation}

The primary prerequisite for imaging flying UAVs is converting their arbitrary translational flight into a stable model.
The core design principle is phase center stabilization, necessitating not only calibrating the spatial positional disparities among distributed ISAC nodes but also compensating for range variations to align echoes across different time instants to a fixed reference point.
A coarse-to-fine calibration scheme can be employed in ISAC networks.
Specifically, the demodulation reference signals in 5G NR can be employed to track the target's coarse phase evolution, providing a robust initialization, significantly narrowing the search space for subsequent fine-tuning.
Subsequently, optimization-driven approaches can be combined with physical kinematic models and data-driven generative priors to enforce sparsity and structural constraints for imaging, ensuring robust convergence and precise phase calibration.

\subsubsection{Rigid-Body Rotation and Aspect Diversity}

While rotational motion generates the Doppler diversity essential for cross-range resolution, it also introduces significant energy migration across adjacent range cells.
Therefore, imaging algorithm design must adhere to two key principles.
First, range-azimuth decoupling: The imaging framework must sever the coupling between the spatial domain and the frequency domain, where the defocusing artifacts caused by rotation should be corrected to ensure that scattering centers remain focused within their respective range cells.
Second, angle-aware holographic fusion: Since the ISAC network observes the target from varying aspect angles, the system must account for the angle-dependent fluctuation of the radar cross section. This requires multi-view joint processing to coherently fuse images across different observation angles, reconstructing stable, holistic morphological profiles.

\subsubsection{Micro-Doppler Interference from Rotor Dynamics}

The high-speed rotation of propellers introduces significant micro-Doppler interference, which violates the rigid-body assumption.
This challenge is exacerbated for ISAC networks using orthogonal frequency division multiplexing signals, where the high-frequency rotation destroys subcarrier orthogonality, inducing severe inter-carrier interference (ICI).
To resolve this, the design principle is signal disentanglement, which separates the non-rigid rotor components from the rigid-body fuselage echoes.
First, the ICI pattern caused by rotor components should be reconstructed and subtracted from the received signals to recover the orthogonality for the remaining fuselage echoes.
Second, DL techniques can be exploited to learn the specific textural features of micro-Doppler interference, ensuring artifact identification and removal.

\subsection{Forensic Imaging of Hovering UAVs}
\label{sec-fine-grained-imaging-2}

This subsection pivots to forensic-level characterization in privacy-intrusion scenarios.
When the ISAC system detects non-cooperative UAVs hovering near a building for an extended duration, it infers a potential risk of illicit video recording.
In this context, the system's objective shifts to immediate alerting and evidence documentation, leveraging the network to capture fine-grained images of the intruder.

The first challenge for hovering UAV imaging is interference management.
On the one hand, it suffers from severe multipath scattering from urban structures and vegetation, which is more pronounced than in flight scenarios.
To address this, enabling techniques including 3D city digital twin and ray tracing can be integrated to generate real-time clutter maps, assisting in precise subtraction of environmental clutter via spatial-temporal filtering.
On the other hand, hovering UAVs also suffer from rotor-induced micro-Doppler interference, which can be effectively separated since the hovering fuselage exhibits negligible Doppler shifts.
Consequently, the fuselage scattered signals serve fine-grained imaging, while the micro-Doppler components reveal kinematic parameters (e.g., blade count, length, and rotation rate), enabling precise UAV identification.

\begin{figure*}[t]
\centering
\includegraphics[width=0.97\linewidth]{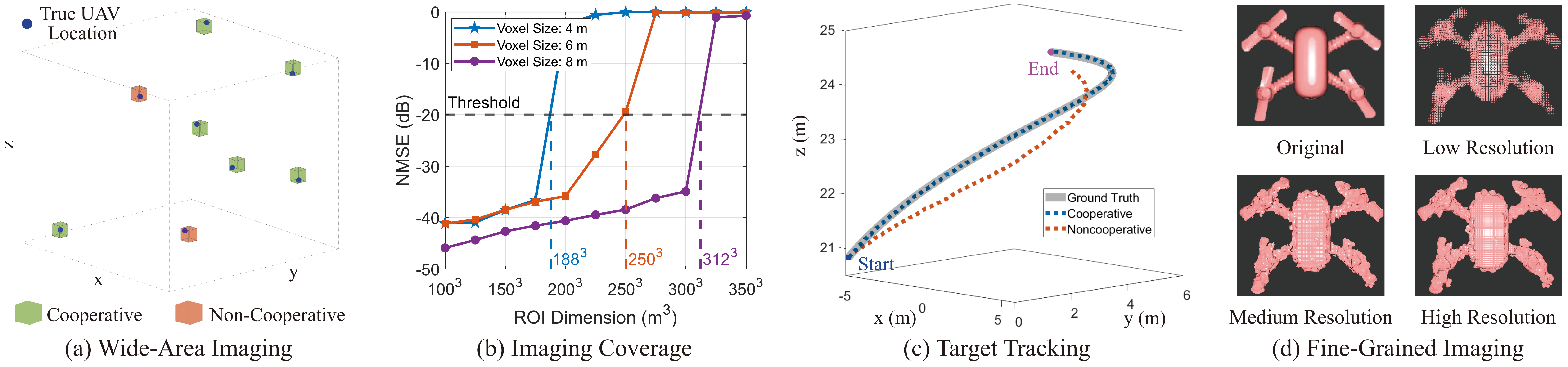}
\caption{Simulation results of the proposed hierarchical imaging framework for low-altitude surveillance.}
\label{fig6}
\end{figure*}

The second challenge lies in imaging precision enhancement.
The objective is to perceive fine-grained target information, necessitating characterizing the target's image as accurately as possible, which is different from point-target sensing discussed in previous sections.
Consequently, emerging DL paradigms such as implicit neural representations (INR) and 3D Gaussian splatting \cite{wu2024embracing} offer novel directions for reconstructing detailed 3D geometries of hovering targets.
Furthermore, this imaging process can also integrate tracking priors and external optical images (captured under proper lighting) to assist in reconstruction.
Ultimately, generative diffusion models can facilitate cross-modal synthesis and generation between radio and optical images, providing multi-dimensional information about the target.

\textbf{Refining the Metric for Target-Centric Imaging:}
Finally, the definition of ``imaging coverage'' should be adapted to target-centric granular perception, evolving into a semantic resolvability zone.
It is defined as a spatial subset of the tracking coverage, within which the ISAC network can harvest sufficient information to reconstruct the target's morphological details, under the constraints of imaging resolution and accuracy.
By quantifying this high-fidelity envelope, the metric provides a critical guideline for network resource scheduling, ensuring a seamless transition from localization to forensic-level identification for proximal threats.

\section{Case Study and Numerical Results}
\label{sec-results}

To evaluate the efficacy of the proposed wireless imaging-based hierarchical surveillance scheme, we conduct an illustrative case study.
The simulation environment features an ISAC network comprising four GBSs \cite{huang2025learned}, which transmit signals into the aerial space and capture echoes, realizing wide-area and target-centric imaging.
The proposed imaging-based framework is validated from the following aspects:

First, we demonstrate wide-area snapshot imaging in Tier I by employing a physics-assisted DL approach, where an intermediate image is first computed using physical models and subsequently refined by an NN.
Fig. \ref{fig6}(a) shows the proposed scheme successfully detects all off-grid UAVs, demonstrating its robustness against grid-mismatch errors that commonly plague compressed-sensing-based imaging.
The imaging results implicitly present the holistic voxel occupancy of the UAVs and their scattering coefficients, providing sufficient information for robust low-altitude monitoring.

Fig. \ref{fig6}(b) further characterizes the imaging coverage under varying resolutions, subject to an accuracy threshold of -20 dB in NMSE.
This threshold can be flexibly adjusted to meet practical requirements on imaging accuracy.
It is observed that the imaging error scales with the ROI dimensions under a fixed resolution.
Excessive expansion of the ROI exacerbates the ill-posed nature of the inverse problem, eventually leading to a complete collapse of reconstruction fidelity.
To maintain the NMSE below -20 dB, the maximum tolerable ROI dimensions at resolutions of 4, 6, and 8 meters are bounded at 188, 250, and 312 meters at each dimension, respectively.
These spatial volume limits define the system's imaging coverage.

Second, we validate UAV trajectory tracking and prediction based on wide-area dynamic imaging in Tier II.
For clarity, we consider one cooperative or non-cooperative UAV.
We utilize the Mamba architecture for multi-frame joint imaging to learn latent motion dynamics and track trajectories.
As depicted in Fig. \ref{fig6}(c), the tracked trajectory of the cooperative UAV closely follows the ground truth, benefiting from the rich, reliable CSI data provided by the communication link.
While the non-cooperative UAV exhibits lower tracking precision due to sparse CSI measurements and lower SNR, the proposed algorithm maintains continuous tracking without target loss.

Third, we perform target-centric fine-grained imaging of an illicitly hovering UAV in Tier III.
Specifically, we employ an INR-based method to learn the target's morphological information, implicitly embedding it within the network parameters.
The imaging results depicted in Fig. \ref{fig6}(d) confirm the successful reconstruction of the UAV's fine-grained features.
Notably, the INR-based method inherently supports arbitrary-resolution imaging, which can generate results tailored to specific resolution requirements.

In summary, these results demonstrate that the proposed imaging-based framework can effectively achieve a robust integration of wide-area monitoring and target-centric characterization for comprehensive low-altitude security.

\section{Conclusion and Future Directions}
\label{sec-conclusion}

This article has explored the potential of wireless imaging for low-altitude surveillance.
Within the proposed hierarchical framework, low-altitude sensing is formulated as snapshot or dynamic, wide-area or target-centric imaging problems, resolved using advanced image reconstruction and processing techniques.
We have discussed how ISAC networks utilize low-altitude imaging schemes to ensure flight safety and foster the burgeoning low-altitude economy.
Sample experiments have validated the effectiveness of the proposed framework.
Several areas could be explored for future advancements:

\textbf{Balanced Communication and Aerial Imaging:}
Determining the optimal allocation of time-frequency resources for low-altitude imaging remains an open question, particularly when simultaneous demands exist for communicating with cooperative users and performing high-dimensional snapshot or dynamic imaging.
In this context, ISAC frame structure design should be further optimized to balance the latency-fidelity trade-off imposed by imaging-centric sensing tasks.

\textbf{ISAC Waveform Design:}
Integrating aerial imaging functions into communication systems requires radio waveforms to not only maximize data transmission rate, but also to support volumetric imaging resolution, spatial coherence, and controllable point-spread functions.
AI-empowered waveform design has been a promising research direction, where large generative models are anticipated to adaptively produce imaging-aware waveforms based on UAV density, CSI quality, channel dynamics, and communication requirements.
 
\textbf{Prototype Validation:}
The developed low-altitude imaging framework should be validated in prototype systems to examine their practical sensing performance.
Specifically, the real-world challenges posed by the low, slow, and small characteristics of UAVs must be rigorously tested.
The experimental results should serve as feedback to refine the imaging algorithm design, thereby facilitating the practical deployment of the proposed framework.

\section{Acknowledgments}

This work was supported in part by the National Natural Science Foundation of China (NSFC) under Grant 62261160576 and Grant 624B2036;
in part by the Fundamental Research Funds for the Central Universities under Grant 2242022k60004;
in part by the National Science Foundation of Jiangsu Province under Grant BK20230818;
in part by the Key Technologies R\&D Program of Jiangsu (Prospective and Key Technologies for Industry) under Grant BE2023022 and Grant BE2023022-1.
The work of C.-K. Wen was supported in part by the National Science and Technology Council of Taiwan under the grant NSTC 114-2218-E-110-006.

\bibliographystyle{IEEEtran}
\bibliography{ref}{}

\small{\noindent{\textbf{Yixuan Huang}} (huangyx@seu.edu.cn) is pursuing the Ph.D. degree in Southeast University, Nanjing, China.\\
{\textbf{Jie Yang}} (yangjie@seu.edu.cn) works with Key Laboratory of Measurement and Control of Complex Systems of Engineering, Ministry of Education, Southeast University, China.\\
{\textbf{Chao-Kai Wen}} (chaokai.wen@mail.nsysu.edu.tw) is a professor with the Institute of Communication Engineering, National Sun Yat-sen University, Kaohsiung, Taiwan.\\
{\textbf{Shi Jin}} (jinshi@seu.edu.cn) is a professor with the School of Information Science and Engineering, Southeast University, China.}

\end{document}